\documentstyle[sprocl,epsfig]{article}

\def\be{\begin{equation}}
\def\ee{\end{equation}}
\def\bea{\begin{eqnarray}}
\def\eea{\end{eqnarray}}

\newcommand{\beal}[1]{\begin{eqnarray}\label{#1}}
\newcommand{\beq}{\begin{equation}}
\newcommand{\eeq}{\end{equation}}
\newcommand{\bel}[1]{\begin{equation}\label{#1}}

\let\o=\omega
\renewcommand{\a}{\alpha}

\renewcommand{\d}{\delta}

\newcommand{\G}{\Gamma}
\newcommand{\g}{\gamma}

\newcommand{\mn}{{\mu\nu}}

\newcommand{\0}{\over }

\newcommand{\2}{\frac{1}{2}}

\newcommand{\6}{\partial}

\def\({\left(}     \def\){\right)}
   
\def\lgk{\left\{}  \def\rgk{\right\}}

\def\gtrsim{\mbox{\,\raisebox{.3ex}{
           $>$}$\!\!\!\!\!$\raisebox{-.9ex}{$\sim$}\,\,}}
\def\7#1{{#1}\llap{/}}
\let\ln=\log

\begin{document}

\begin{flushright}
\tt hep-ph/9707538\\
TUW 97-14\\
\end{flushright}
\vspace{0.4cm}

\title{IMPROVED HARD-THERMAL-LOOP EFFECTIVE 
ACTIONS \footnote{Invited talk given
at SEWM '97, May 22--25, 1997, Eger, Hungary.}
}

\author{ A. REBHAN }

\address{Institut f. Theoretische Physik, Tech. Univ. Wien\\
Wiedner Hauptstr. 8--10, A-1040 Vienna, Austria}


\maketitle\abstracts{
Hard thermal loop effective actions furnish the building blocks of
resummed thermal perturbation theory, which is expected to work
as long as the quantities under consideration are not sensitive to
the nonperturbative (chromo-)magnetostatic sector.
A different breakdown of perturbation theory occurs whenever
external momenta are light-like, because the hard thermal loops
themselves develop collinear singularities. By additionally 
resumming asymptotic
thermal masses for hard modes these singularities are removed while
leaving the gauge invariance of the effective action intact.
}
 
\section{Review of Hard-Thermal-Loop Resummation}

Precisely at (and below) those energy/momentum scales where perturbation theory
of quantum field theories at non-zero temperature\cite{appl} 
predicts new effects
from collective behaviour like screening, plasma frequencies, Landau
damping etc., the conventional loop expansion ceases to be reliable.
The infrared singularity of the Bose-Einstein distribution
\be
n_B(k)={1\0 e^{k/T}-1} \sim {T\0k} \quad
\mbox{for $k\to0$}
\ee
causes an increasing sensitivity to the infrared as the loop order
increases. Eventually, this gives rise to infrared
divergences which inhibit further refinements of the results or which
even may prevent one from determining the leading terms.

However, the very same nontrivial screening masses, plasma frequencies,
and Landau damping effects in most cases provide the remedy for
perturbation theory, because once taken into account they provide a
cut-off to otherwise infrared divergent loop integrals.

One of the simplest cases where one can observe this mechanism of self-healing
is a massless scalar field theory with quartic self-interactions $g^2\phi^4$.
At finite temperature, one finds for the thermal contribution to
the one-loop scalar self-energy a simple constant proportional to $T^2$,
\be
\Pi\big|_{T>0}-\Pi\big|_{T=0}=g^2T^2=m_{\mathrm th}^2
\ee
which can be interpreted as a thermal mass for the scalars in
a heat bath. It is not a mass in the usual sense, though. For instance
it does not contribute to the trace of the 
energy-momentum tensor,\cite{NRS} which
a mass usually does.
But it does modify the infrared behaviour of scalar particles
by introducing a mass gap to their dispersion laws.
Clearly, this mass term becomes important for momenta of the order $gT$.
At and below this scale, it is at least as important as the usual
kinetic term, and the effective propagator becomes 
$[k_0^2-\vec k^2-m_{\mathrm th}^2]^{-1}$ instead of the massless one.
As a consequence, infrared divergences in higher loop diagrams become
cut off at the scale of $m_{\mathrm th}$ and lead to an effective
loop expansion parameter $g^2T/m_{\mathrm th}\sim g$.

Including the thermal mass into the scalar propagator corresponds
to a reorganization of perturbation theory. Overcounting is
avoided by writing
\be
{\cal L}=\underbrace{{\cal L}-{1\02}m_{\mathrm th}^2 \phi^2}_{{\cal L}_{\mathrm
eff}}+\underbrace{{1\02}m_{\mathrm th}^2 \phi^2}_{{\cal L}_{\mathrm counter}}
\ee
and treating the thermal mass term as both a contribution to the
classical action and as a one-loop counter-term.
Alternatively\cite{Breff}, one may view ${\cal L}_{\mathrm
eff}$ as (the leading terms of)
the effective low-energy action obtained after integrating
out the hard modes with momenta $\!\!\gtrsim T$. These are indeed solely
responsible for the
appearance of a thermal mass for soft modes.

Turning to gauge theories now, a well-known collective phenomenon at
finite temperature or density is Debye screening of
electric fields. In a static situation
one finds an effective Lagrangian with a mass term for the zero components
of the vector potential,
\be
{\cal L}_{\mathrm eff}=-{1\02}m_D^2 A_0^2
\ee
with $m_D^2={1\03}e^2T^2$ in the high-temperature limit. The necessity
for a resummation of this mass term has been noted in the context of QED
first by Gell-Mann and Brueckner\cite{GMB}. Otherwise infra-red divergences
appear at three-loop order of a perturbative evaluation
of the partition function.

In nonabelian gauge theories like QCD there is similarly a screening mass
for chromo-electrostatic modes. The leading-order term differs only
by the replacement of $e^2$ by $g^2(N+N_f)$ for gauge group SU($N$) and
$N_f$ fermions. However, because of the self-interactions of the
gluons, Debye screening is not sufficient for curing the infrared
problem of perturbation theory. The (chromo)magnetostatic modes
are unscreened 
at the scale $gT$
and are therefore expected to cause
divergences beyond three loop order in the partition function.
If an infrared cut-off effectively arises at the next lower scale
$g^2T$, this will not cure perturbation theory as it did above:
the effective expansion parameter $g^2T/m_{\mathrm th}$ is then
of order 1. In what follows we shall always be concerned only
with that part of (resummed) perturbation theory where one is still
below those loop orders where this presumably insurmountable
barrier to perturbation theory appears.

If one is interested in computing thermal effects in nonstatic
quantities, it turns out that taking into account Debye screening
is not sufficient. 
The leading contribution to the gauge bosons' self-energy
is a nontrivial function of the temporal and spatial components of
the momentum vector involving Legendre functions of the second kind
and so represents
a nonlocal contribution to an effective Lagrangian. Moreover
the loss of Lorentz invariance (through the distinguished rest frame
of the plasma) leads to two independent structure functions
in the polarization tensor,
\be
\Pi^\mn(q_0,q)=\Pi_t(q_0,q)P_t^\mn+\Pi_\ell(q_0,q)P_\ell^\mn
\ee
where $P_{t,\ell}^\mn$ are two symmetric tensors that are transverse
with respect to the 4-momentum $Q=(q_0,\vec q)$. One can be chosen
to be transverse also with respect to 3-momentum and will be denoted as
$P_t$; the second, $P_\ell$, will be taken to be orthogonal to the former.

In nonabelian gauge theories, $\Pi^\mn$ does not necessarily have to
be transverse, but one can always find gauges where it does. Assuming
transversality, the functions $\Pi_{t,\ell}$
are given by
\be
\Pi_\ell=-{Q^2\0q^2}\Pi_{00}, \quad \Pi_t={1\02}(\Pi^\mu{}_\mu-\Pi_\ell)
\ee
in the rest frame of the plasma.
The leading terms in the high-temperature expansion of the one-loop
diagrams for the self-energy are relevant for the infrared domain
$\o,q \ll T$ and are given explicitly by\cite{Silin}
\bea
\Pi^\mu{}_\mu&=&{e^2T^2\03},\\
\Pi_{00}&=&{e^2T^2\03}\(1-{\o\02q}\ln{\o+q\0\o-q}\),
\label{Pi00}
\eea
where in the imaginary-time formalism an analytic continuation
$Q_0=2\pi n T \to \o+i\epsilon$ has been made. In the case of
QCD the result, to leading order, is the same up to the
substitution $e^2\to g^2(N+N_f)$.\cite{KKW}

 \begin{figure}                                            
     \centerline{ \epsfxsize=4.5in                           
     \epsfbox[85 310 500 485]{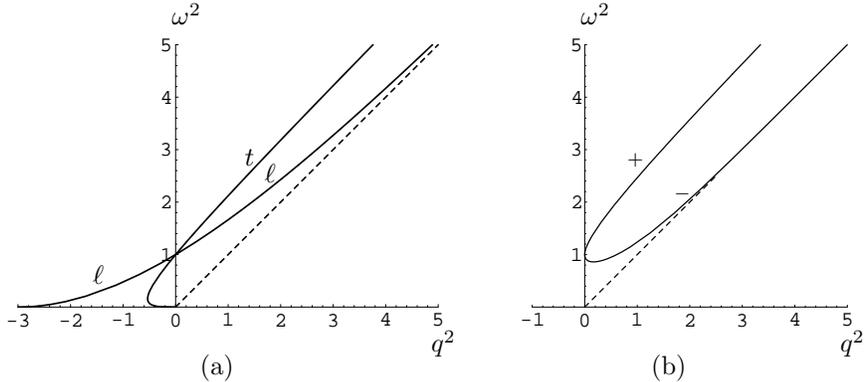} }                    
 \vspace{-4.5cm}  \font\frm=cmr10
 \unitlength0.75cm \begin{picture}(16,5)
 \put(3.5,5.0){$\omega^2$} \put(10.75,5.0){$\omega^2$}
 \put(11.57,2.6){$_+$}    \put(12.4,2){$_-$}
 \put(4.8,2.5){$t$}       \put(5.15,2.2){$\ell$} 
 \put(2.1,.4){$\ell$}     \put(8.1,-.8){$q^2$}  
 \put(15.35,-.8){$q^2$}  
 \put(4,-1.2){(a)} \put(12,-1.2){(b)}
\end{picture} \\[20pt] 
\caption{
The leading-order gluonic (a) and fermionic (b) dispersion laws. For the
transverse ($t$) and longitudinal ($\ell$) dispersion curves
of the gluonic modes $\omega^2$ and $q^2$ are given in units
of $m^2$; for the quark ($+$) and plasmino ($-$) modes the
unit is $M^2$.}
\end{figure}                                               

Corresponding to the two components in the self-energy tensor, there
are now two different ``mass-shells'' in the (Dyson resummed)
photon/gluon propagator, which are shown in fig.~1a.

Since quadratic scales have been used in this figure, a Lorentz invariant
mass shell would be given by a straight line parallel to the light-cone
which is represented by a dashed line. The full dispersion law can
no longer be given in closed form but
has to be obtained by numerically solving for poles in the propagator.
For positive $q^2$, one finds
propagating modes above the plasma frequency $\o_{\mathrm pl}=m\equiv
eT/3$. The spatially transverse branch, which corresponds
to radiative modes, approaches asymptotically
an ordinary mass-shell with mass $m_\infty=eT/\sqrt6$.
The longitudinal branch describes a collective mode which has no
analogue in the vacuum. Its residue decays exponentially fast with
increasing $q$, for which it approaches the light-cone.

For frequencies below $\o_{\mathrm pl}$, the only poles in the propagator
occur for imaginary values of $q$ (negative values of $q^2$ in
fig.~1a) which correspond to screening effects.
The two branches show different screening lengths, and in the static
limit only the longitudinal one has a finite (Debye) screening length,
whereas the transverse branch has a vanishing static screening mass.

Another feature worth mentioning is that the polarization tensor has
a large imaginary part for space-like momenta. Its physical interpretation
is given by
the effect of Landau damping, which corresponds to the possibility
of absorption of soft modes by the hard constituents of the plasma.

In fig.~1b, the dispersion laws for fermionic matter is given. Again
one finds a second, collective mode, which technically arises
because the loss of Lorentz invariance leads to two independent
structure functions in the fermion self-energy. In the
high-temperature limit these are determined by\cite{KW}
\bea
{1\04}{\mathrm tr}\; \7Q \Sigma &=& {e^2T^2\08} \\
{1\04}{\mathrm tr}\; \g^0 \Sigma &=& {e^2T^2\016q}\ln{\o+q\0\o-q} \;.
\eea
(In the case of QCD, the result differs only by the replacement
$e^2 \to g^2{N^2-1\02N}$, as far as the leading terms given here
are concerned.)

The resummed fermion propagator has the form
\be
S_F(\o,q)=\2\(\g_0+{\7q\0q}\)D_+(\o,q)+\2\(\g_0-{\7q\0q}\)D_-(\o,q) \;.
\ee
Both branches have a common frequency $\omega=M=eT/2$ at $q=0$.
The ($+$)-branch tends asymptotically to $M_\infty=\sqrt2 M$. The
($-$)-branch on the other hand, which is sometimes termed
``plasmino'',\cite{Pispha} approaches the light-cone with
exponentially vanishing residue, pointing at its nature of
a collective mode. It has a curious dip in its dispersion law
which is reminiscent of rotons in superfluidity. It is
anomalous also in that it has a reversed sign in the ratio of
helicity and chirality, which comes from the fact that it
describes the behaviour of hole (rather than particle)
excitations in the plasma.\cite{Weldon}

Notice that contrary to the bosonic case, there are no poles in the propagator
for $q^2<0$, corresponding to the impossibility of a screening of fermion 
numbers. But as in the bosonic case, there is Landau damping
for space-like momenta, since soft fermions can be annihilated by
hard ones or can be transformed into hard ones by absorption of
hard gauge bosons.

Evidently, the hard-thermal-loop modifications of the basic propagators
are such that they can no longer be described by a simple effective mass term.
The corresponding effective action is instead a highly nonlocal object,
and it is moreover nonpolynomial in the fields, because there are
also (nonlocal) vertex corrections that have to be treated on a par
with the tree-level vertices once we are considering momenta of the
order $eT$. It turns out that all one-loop diagrams that have
contributions proportional to $T^2$ in the high-temperature limit
are of the same order of magnitude as tree-level vertices when
$\o,q\sim eT$. Such hard-thermal-loop vertices exist for two fermions
together with any number of gauge bosons and, in the nonabelian case,
for any number of gluons in interaction.\cite{HTL} They happen to vanish
in the static limit, where the effective action becomes local.

Remarkably, this nonlocal, nonpolynomial effective action can be
given in a rather compact form involving an additional
integration over the directions of an auxiliary light-like vector.
Taylor and Wong\cite{TW} have shown that this effective action
can be written as
\bel{EATW}
\G[A]=-m_D^2\( \2\int d^4x A_0^a A_0^a + \int{d\Omega_{\vec p}\04\pi}
W(A\cdot P) \)
\ee
where
\be
W(A\cdot P)={\mathrm tr}\;\int d^4x \6_0(A\cdot P) f({1\0P\cdot\6}[iA\cdot P,*])
{1\0P\cdot\6}A\cdot P
\ee
with $f(z)=2\sum_0^\infty{z^n\0n+2}$ and $P=(1,\vec p)$, $P^2=0$.
The average over the latter is the remnant of having integrated out
the hard massless modes of the plasma constituents.

An alternative form which makes the gauge invariance of this effective
action manifest has been given by Braaten and Pisarski and reads\cite{BPeff}
\beal{EABP}
\G[A]&=&-\4m_D^2\int{d\Omega_{\vec p}\04\pi}\int d^4x F^{\mu\sigma}_a
{P_\sigma P_\rho\0[ P\cdot D(A) ]_{ab}^2} F^\rho_{b\mu} \nonumber\\
&&+iM^2\int{d\Omega_{\vec p}\04\pi}\int d^4x \bar\psi {\7P\0P\cdot D(A)}\psi
\eea
where we have now included also the fermionic part.

Hard-thermal-loop resummation means that one has to use
resummed propagators and resummed nonlocal vertices in loop calculations.
For example, the calculation of the
production rate of virtual soft photons (soft dileptons)\cite{BPYW}
involves evaluating the imaginary part of the diagrams given in fig.~2.
Its building blocks can be cut in a number of different ways.
Cutting a soft line gives contributions from the quasi-particle poles
as well as from Landau damping, which is given by the cut of the
self-energy. Moreover, hard-thermal-loop vertices have a number of
different cuts,
 as depicted
in fig.~3. In these figures,
hard and soft momenta are indicated by thin and heavy lines, respectively.

\begin{figure}[b]                                            
     \centerline{ \epsfxsize=3.5in                           
     \epsfbox[80 400 410 485]{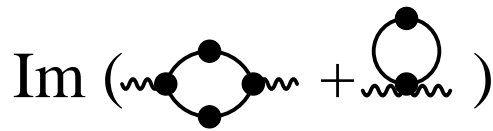} }                    
\caption{One-loop resummed contributions to soft photon production}
\end{figure}

\begin{figure}[t]                                            
     \centerline{ \epsfxsize=3.5in                           
     \epsfbox[140 400 495 605]{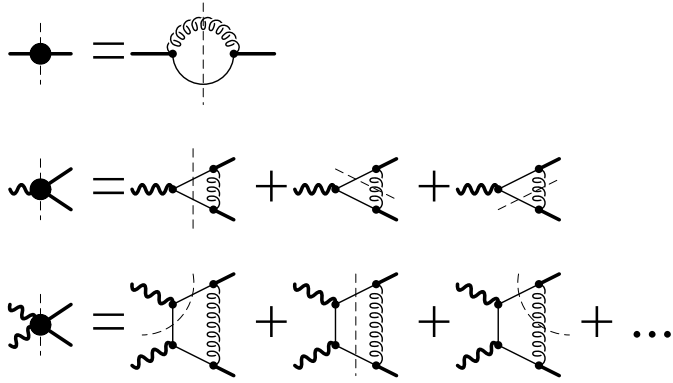} }                    
\caption{Various contributions from cutting resummed quantities}
\end{figure}

\begin{figure}[b]                                            
     \centerline{ \epsfxsize=3.5in                           
     \epsfbox[80 400 410 460]{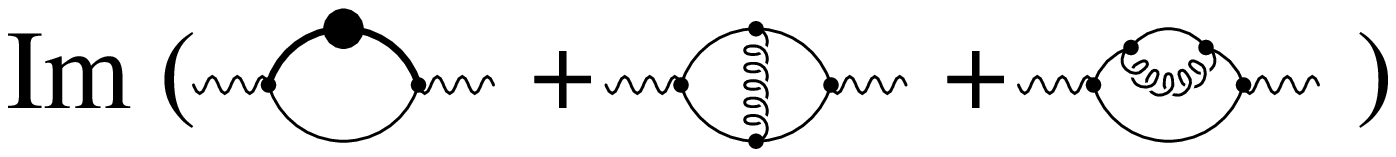} }                    
\caption{One-loop resummed contributions to hard photon production}
\end{figure}

This approach has been applied successfully to a number of other
problems\cite{appl}, too, and even in those cases where the quantities
under consideration turns out to be sensitive to the magnetic mass scale,
it is usually possible to extract a leading logarithmic correction.

A quite different difficulty has been encountered recently in attempts
to calculate the production rate of soft real photons from a quark-gluon
plasma\cite{BPS}.

\section{Collinear singularities}

It seems to make sense physically to take the thermodynamic large-volume limit
only with respect to strongly interacting matter and to assume that
particles 
without strong interactions can escape immediately.

This is the rationale behind the above-mentioned calculation of
the production rate of virtual photons which decay into dilepton pairs.
Similarly one may consider real photons which are produced by the
electrically charged quarks within the plasma.

Technically the simplest case is the one of hard real photons.
Because of the external hard momentum that has to be routed through
the diagrams, there is at most one soft propagator involved that
needs to be dressed. In particular, there are no complicated hard-thermal-loop
vertices to be taken into account.
The relevant diagrams are shown in fig.~4.

This calculation has been performed in Ref.~\citelow{KLSBNNR} 
with the result,
when the energy of the photon $E\sim T$,
\be
E{dW^{\mathrm HRP}\0d^3q} \propto e^2 g^2 T^2 \ln{cT\0gT} \;,
\ee
where $c$ is a calculable constant.
Here resummation has replaced a log of the bare quark mass by
$\ln(gT)$, thereby screening the collinear singularity that would
appear in the limit of vanishing (or negligible) bare quark masses.

The analogous calculation for {\em soft} real photon production instead
involves the diagrams of fig.~2, but in this case it turns out that
the resummed result still diverges for
massless quarks as the light-cone is approached,
\be
E{dW^{\mathrm SRP}\0d^3q} \propto e^2 g^3 T^2 \ln{c'T\0gT} \ln{c''T^2\0Q^2}\;.
\ee

The by now standard hard-thermal-loop resummation method appears to
break down here, but in a novel way, for the magnetostatic sector is
innocuous this time. The difficulties are rather rooted in the
hard thermal loops themselves. They, too, exhibit collinear singularities
when one of the external soft momenta approaches the light-cone.

For example $\Pi_{00}$ as given in (\ref{Pi00}) 
diverges logarithmically when $|\o|\to q$.
Before carrying out the integration over hard momenta $\Pi_{00}$ reads
\bel{Pi00int}
\Pi_{00}(Q) = -2e^2 \int\!{d^3p \0 (2\pi)^3}\,n'(p)
\lgk 1-{Q_0 \0
 Q_0-\vec p\vec q/p} \rgk
\label{poohtl}\ee
so this indeed comes from a collinear singularity, which for light-like
$Q$ arises when
the soft momentum $\vec q$ becomes parallel to the hard momentum $\vec p$.


This singularity of $\Pi_{00}$ looks harmless inasmuch as the
longitudinal plasmon mass-shell, which is determined by  $\Pi_{00}$,
never reaches the light-cone, although it does
approach it exponentially. However, a problem similar to the
one in the soft-photon production rate 
appears if one wants to calculate next-to-leading order
corrections to the plasmon dispersion law in the vicinity of the
light-cone. In Ref.~\cite{KRS} this has been done in the case
of scalar electrodynamics with the result that
\beq
\delta \Pi_{00}(Q) \sim {em^2q\0\sqrt{Q^2}}
\label{dPi} \;.
\eeq
This is even more singular than the leading, hard-thermal-loop result,
so sufficiently close to the light-cone, namely when
$Q^2/q^2 \ll e^2$, we have that
$\delta \Pi_{00}\gg \Pi^{\mathrm HTL}_{00}$,
and the resummed perturbation theory clearly is breaking down.

It is precisely the singularity of $\Pi_{00}$ at the light-cone that
is causing this trouble here, and this singularity is possible only
by the masslessness of the hard modes. However, hard modes also acquire
thermal masses through interactions with the plasma constituents. Usually,
they are completely negligible, because they are suppressed by powers
of the coupling constant when compared to the hard momentum scale.
But this is no longer true in the vicinity of the light-cone.
Indeed, including the thermal mass of the scalar particles renders
$\Pi_{00}$ regular, and leads to
\beal{Pi00reg}
\Pi_{00}^{\mathrm leading}(Q^2=0) &=& {e^2\0\pi^2}
\int_0^\infty dp\,p\,n(p)\(1-2\ln{2p\0m_{\mathrm th}}\) \nonumber\\
&=& -{e^2T^2\03}\(\ln{2T\0m_{\mathrm th}}+{1\02}-\g+{\zeta'(2)\0\zeta(2)}\) \;.
\eea

Computing now the next-to-leading order term gives
\be
\d\Pi_{00}(Q^2=0)={e^2Tm_{\mathrm th}\02\pi}
\ee
which because of $m_{\mathrm th}\propto eT$ is smaller than the
leading order result (\ref{Pi00reg}), as required for a sensible
perturbative scheme.

With $\Pi_{00}$ being regular at $Q^2=0$, it turns out that there is
a qualitative change of the plasmon dispersion law. Instead of being
asymptotic to the light-cone, it now pierces the light-cone at
a finite value of $q^2=q^2_c\sim{1\03}e^2T^2\ln(1/e)$.
The failure of (standard) hard-thermal-loop resummation is just a
reflection of the incapability of any perturbation theory to
describe qualitative changes. Hence the necessity to modify the
starting point, namely, the hard thermal loops themselves.

There is in fact nothing wrong with the existence of space-like plasmon
modes: their group velocity remains smaller than 1 throughout.
A similar phenomenon has already been found in 1961 by Tsytovich\cite{Ts}
in the case of non-ultrarelativistic quantum electrodynamics (i.e.,
with temperatures $T<m_e$).

Actually, the importance of these modes is reduced by the corresponding
residue having dropped significantly in the vicinity of the light-cone.
Moreover,
for space-like momenta, strong Landau damping sets in, quickly
rendering them over-damped.

The on-set of Landau damping is usually discontinuous. With thermal
masses for the hard modes taken into account, this is in fact smoothed out
as well. The imaginary part of $\Pi_{00}$ sets in as
\be
{\mathrm Im}\,\Pi_{00}(Q^2<0)\Big|_{q-\o\ll e^2q}=
{e^4T^2q\08\pi(q-\o)}e^{-e\sqrt{q\08(q-\o)}}
\ee
that is, starting from zero with all its derivatives vanishing.

\section{Improved hard-thermal-loop effective action}

In the above example of scalar electrodynamics, the collinear singularities
within the hard thermal loops were removed by a resummation of
the thermal mass of scalar particles for both soft and hard
modes. In more complicated gauge theories we have seen that 
one is usually confronted
with more than one dispersion law for a given species. However, it turns
out that additional collective modes such as the longitudinal plasmon
and the plasmino do not contribute to the hard thermal loops. Whereas
the additional modes 
would have zero mass asymptotically, the
transverse vector boson branch and the ($+$)-branch of the fermionic
modes both tend to a mass hyperboloid with asmyptotic masses $m_\infty$
and $M_\infty$, respectively. Therefore we can expect that the
collinear singularities in hard thermal loops on the light-cone
will again be cut off once these asymptotic masses are taken into account.

Our improved hard thermal loops\cite{FR} 
can be obtained almost exactly as usually
with the only difference that the (hard) gluon propagator reads
\be
D^{\mu\nu}(P)={1\0P^2-m_\infty^2}P_t^{\mu\nu}+{1\0P^2}(P_\ell^{\mu\nu}+
\alpha {P^\mu P^\nu\0P^2})
\ee
where we have used a general covariant gauge, and the fermion propagator
\be
S_F(P)={\7P\0P^2-M_\infty^2} \;.
\ee

In (\ref{Pi00int}) the expression $\vec p\vec q/p$ comes from
the first term in the expansion of the energy difference $\o(\vec p)-
\o(\vec p-\vec q)$ for $|\vec q|\ll|\vec p|$. However, for light-like $Q$
and collinear 3-momenta, the next term in the expansion becomes important,
and we have $\o(\vec p)-
\o(\vec p-\vec q) \approx \vec p\vec q/p - \vec p\vec q m_\infty^2/2p^3$
if the hard modes involved are gluons (when the loop diagram is made from
quarks, one just has to replace $m_\infty$ by $M_\infty$).

Since this extra contribution is only needed in the vicinity of the light-cone
and when the 3-momenta are close to collinearity, the same regularization
is obtained by replacing $Q_0 \to (1+m_\infty^2/2p^2)Q_0$ in the
critical denominators. Because the potential collinear singularities
are only logarithmic, the numerators can be simplified as usually
in hard thermal loops. This holds true also for vertex functions as long
as exceptional momentum configurations are avoided.

As a consequence, we find that the structure of the hard-thermal-loop
effective action is largely preserved. In the version of Taylor and Wong
(\ref{EATW}) we have to modify the averaging over the directions
of $\vec p$ according to
\beal{modTW}
\int {d\Omega_{\vec p}\04\pi} f(P) &\to&
{3\0\pi^2}\int_0^\infty d\a{\a^2 e^\a\0(e^\a-1)^2}
\int {d\Omega_{\vec p}\04\pi} f(P(\a))\nonumber\\
P=(1,\vec p) &\to& P(\a)=(\sqrt{1+{m_\infty^2/T^2\0\a^2}},\vec p)
\eea
with $\vec p^2=1$.
These modifications indeed do not spoil any of the properties that
are crucial for the gauge invariance of the effective action.

A similar modification applies for the manifestly
gauge invariant form of the effective action due to Braaten and Pisarski
(\ref{EABP}). It differs from (\ref{modTW}) in that
\be
{3\0\pi^2}\int_0^\infty d\a{\a^2 e^\a\0(e^\a-1)^2}\cdots \to
{6\0\pi^2}\int_0^\infty d\a{\a e^\a\0e^\a-1}\cdots
\ee
and furthermore in that $P_0(\a)$ can no longer be simplified to 1
when it appears in numerators. This is because in the form (\ref{EABP})
there are individual contributions which prior to the modification
had collinear singularities that were stronger divergent than only
logarithmically, and so more care has to be exerted when working with
the manifestly gauge invariant version.

Another complication occurs when external fermion lines are involved.
Then two different massive propagators occur inside hard thermal loops
and the dominant term that shifts the pole causing collinear singularities
is proportional to the difference of the bosonic and fermionic asymptotic
masses and is independent of the soft momentum. The modification to
be applied to the hard-thermal-loop effective action is then
\be
\int {d\Omega_{\vec p}\04\pi}{\cdots\0iP\cdot D(A)} \to
{2\0\pi^2}\int_0^\infty d\a{\a e^\a\0e^{2\a}-1}
\int {d\Omega_{\vec p}\04\pi}\left[{\cdots\0iP\cdot D(A)
+{m_\infty^2-M_\infty^2\02\a T}} - {\mathrm c.c.} \right]
\ee
where $P$ remains a null vector, $P^2=0$.
The manifest gauge invariance of the fermionic 
hard-thermal-loop effective action is clearly
left intact by this replacement.

\section{Open problems}

Although the collinear singularities are removed by a resummation of
asymptotic thermal masses for the hard modes that yield the
hard-thermal-loop effective action, it is not clear that the
correspondingly improved hard-thermal-loop effective action
leads to a resummed perturbation theory where higher loop orders
are suppressed by higher powers of the coupling.

\begin{figure}[t]                                            
     \centerline{ \epsfxsize=3.5in                           
     \epsfbox[145 400 475 465]{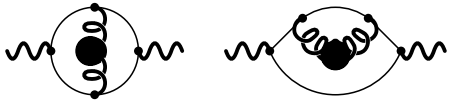} }                    
\caption{Enhanced two-loop contributions to soft photon production}
\end{figure}

In fact, in Ref.~\citelow{AGKP} it has recently been shown that
in the case of the soft-real-photon production rate certain
resummed two-loop diagrams involving both hard and soft internal
momenta (shown in 
fig.~5) are even {\em enhanced} by a factor $1/g^2$ compared to the
contributions of soft one-loop diagrams.

This need not signal a complete breakdown of perturbation theory, for
the one-loop diagrams are unnaturally small because of a statistical
suppression factor. But it certainly raises the question whether
there could be even more enhanced higher-loop diagrams from
stronger and stronger would-be collinear singularities.

If this is the case, one would probably require additional resummations
beyond those of asymptotic thermal masses. A candidate\cite{Nie} would be
the anomalously strong damping of hard modes from interactions with
soft ones,\cite{LS,damp} whose damping constant $\g$ 
is of the order $g^2T\ln(1/g)$.

Although formally of higher order than asymptotic thermal masses,
such a damping term would indeed be more effective in cutting off the collinear
singularities, since
\beq
{Q^2\0q^2}\to {Q_0\g\0q^2}\sim g \gg {m_\infty^2\0T^2} \sim g^2.
\eeq
(The damping contributed by purely hard self-interactions on the
other hand is of the order
$\g\sim g^4T$ and so is less important than the asymtotic thermal mass.)

But in contrast to a mass resummation
it is generally not sufficient to modify only the propagators, when
damping is to be taken into account. Including damping only into the
propagators violates the Ward identities. In Abelian theories the latter
imply $\Pi_{\mu\nu}Q^\nu\equiv 0$, but one finds e.g.
\beq
\Pi_{0\nu}(Q_0,0)Q^\nu\Big|_{\vec q=0}={2i\g Q_0\0Q_0+2i\g}{e^2T^2\03}.
\eeq
Correcting also the vertices 
restores gauge invariance, but it turns out that $\g$ gets eliminated
not only from $\Pi_{\mu\nu}Q^\nu$, but from all the components of
$\Pi_{\mu\nu}$!\cite{LS,KRS}

Moreover, damping effects from interactions
with
soft modes cannot be encorporated into an effective action
which summarizes the effects of
integrating out the hard modes only. 
It is therefore much more difficult to find a consistent
scheme. 
Instead, one would
have to solve the hierarchy of Schwinger-Dyson equations.

Because of all these difficulties, the status of hard-thermal-loop
resummation for processes involving light-like momenta is still
rather unclear and remains a challenging problem.\cite{FRprep}


\section*{Acknowledgments}

This work has been done in collaboration with F. Flechsig and H. Schulz
(Univ.\ Hannover), and U. Kraemmer (TU Wien). 
It has been
supported in part by the Austrian ``Fonds zur F\"orderung der wissenschaftlichen
Forschung (FWF)'', project no. P10063-PHY and by
the Jubil\"aumsfonds der \"Osterreichischen
Nationalbank, project no.~5986.
Moreover I would like to thank the organizers of this workshop for
all their efforts.

\end{document}